\documentclass[3p,times,twocolumn]{elsarticle}

\usepackage{amssymb}

\biboptions{sort&compress}

\usepackage[figuresright]{rotating}

\usepackage{slashbox}
\usepackage{graphicx}
\usepackage{dcolumn}
\usepackage{bm}
\usepackage{hyperref}
\usepackage[mathlines]{lineno}
\usepackage{blkarray}
\usepackage{mathtools}
\usepackage{bbold}
\usepackage{epsfig}
\usepackage{subfig}
\newcolumntype{K}[1]{>{\centering\arraybackslash}p{#1}}
\usepackage{xcolor}
\begin{document}
\def\one{{\mathchoice {\rm 1\mskip-4mu l} {\rm 1\mskip-4mu l} {\rm
\mskip-4.5mu l} {\rm 1\mskip-5mu l}}}
\begin{frontmatter}




\title{Randomized benchmarking of quantum gates implemented by electron spin resonance}


\author[a,b]{Daniel K. Park$^{*,}$}
\author[a,b]{Guanru Feng$^{*,}$}
\author[a,b]{Robabeh Rahimi}
\author[a,b,c]{Jonathan Baugh}
\author[a,b,d,e]{Raymond Laflamme}

\cortext[f]{These authors contributed equally to this work.}

\address[a]{Institute for Quantum Computing, Waterloo, Ontario, N2L 3G1, Canada}
\address[b]{Department of Physics and Astronomy, University of Waterloo, Waterloo, Ontario, N2L 3G1, Canada}
\address[c]{Department of Chemistry, University of Waterloo, Waterloo, Ontario, N2L 3G1, Canada}
\address[d]{Perimeter Institute for Theoretical Physics, Waterloo, Ontario, N2J 2W9, Canada}
\address[e]{Canadian Institute for Advanced Research, Toronto, Ontario M5G 1Z8, Canada}

\begin{abstract}

Spin systems controlled and probed by magnetic resonance have been valuable for testing the ideas of quantum control and quantum error correction. This paper introduces an X-band pulsed electron spin resonance spectrometer designed for high-fidelity coherent control of electron spins, including a loop-gap resonator for sub-millimeter sized samples with a control bandwidth $\sim 40$ MHz. Universal control is achieved by a single-sideband upconversion technique with an I-Q modulator and a 1.2 GS/s arbitrary waveform generator. A single qubit randomized benchmarking protocol quantifies the average errors of Clifford gates implemented by simple Gaussian pulses, using a sample of gamma-irradiated quartz. Improvements in unitary gate fidelity are achieved through phase transient correction and hardware optimization. A preparation pulse sequence that selects spin packets in a narrowed distribution of static fields confirms that inhomogeneous dephasing ($1/T^*_2$) is the dominant source of gate error. The best average fidelity over the Clifford gates obtained here is $99.2\%$, which serves as a benchmark to compare with other technologies. 

\end{abstract}


\end{frontmatter}


\section{Introduction}
\label{sec:intro}

Pulsed electron spin resonance (ESR) techniques have been developed and applied in many areas, including physics, chemistry and biochemistry \cite{distance8,distance9,distance10}. Another important application is quantum information processing (QIP) \cite{CPMGPRA,CPMGPRL,sbus,khaneja2007PRA,hodges2008universal,zhang2011coherent,wuhua,HBAC_daniel}. Systems in which an electron is coupled to nuclei through the hyperfine interaction are advantageous over conventional nuclear magnetic resonance (NMR) QIP experiments. The electron spin polarization can be transferred to the nuclear spins \cite{HBAC_daniel,maly2008dynamic,barnes2008high,simmons2011entanglement,filidou2012ultrafast}, and fast gate operations can be realized using only microwaves, if the hyperfine interaction has a strong anisotropic component \cite{khaneja2007PRA,hodges2008universal,zhang2011coherent}. While gate operations are done through the electron via microwave control, relatively long coherence times of the nuclei can be exploited for storing information \cite{TroyPRL}. Thus the electron-nuclear hybrid spin system is a promising for developing advanced QIP architectures ~\cite{sbus,TroyPRL,TroyThesis}. 
Such systems permit testing the ideas of quantum control and quantum error correction \cite{knill1997theory,knill1998resilient,preskill1998reliable,knill2005quantum,aliferis2007accuracy,gottesman1997stabilizer} in a setting unavailable to classical simulations, opening a path to the development of large scale quantum devices based on spins. 

The ability to generate accurate control pulses with arbitrary amplitudes and phases is required to realize a universal set of quantum gates with high fidelity. Moreover, in other spectroscopic techniques such as double electron-electron resonance (DEER) \cite{distance2,DEER2} and solution-state 2D electron-electron double resonance (2D-ELDOR) \cite{distance4, distance8, Elder}, arbitrary waveform generation is beneficial for creating desired excitation profiles \cite{AWG1,AWG2,AWG3,AWG4}. Integrating an arbitrary waveform generator (AWG) into pulsed ESR spectrometers has been reported in several previous works \cite{hodges2008universal,zhang2011coherent,AWG1,AWG2,AWG3,AWG4,AWG5,AWG6}. In order to achieve precise coherent control, efforts have also been made to overcome pulse distortions due to limited bandwidth of the resonator and amplifier non-linearity \cite{zhang2011coherent,AWG1,AWG2,AWG3,AWG4,Troyringdown}. However, quantitative characterization of unitary gate fidelities and detailed studies of the sources of infidelity in pulsed ESR systems have been rarely addressed \cite{CPMGPRA,CPMGPRL,AWG1}.  

Randomized benchmarking (RB)~\cite{emersonRBM,PhysRevA.77.012307,dankert2009exact, PhysRevLett.106.180504,PhysRevA.85.042311} is a well-developed, scalable approach for estimating the average error probabilities of quantum gates apart from the state preparation and measurement (SPAM) errors. When comparing to the CP/CPMG method for quantifying single-qubit rotation amplitude and phase control errors applied in refs. \cite{CPMGPRA, CPMGPRL}, RB has the advantage of being able to assess a broader class of operations than only $\pi$ pulses. In the work of ref. \cite{AWG1}, a method was proposed to characterize the control accuracy by measuring the excitation profiles of different pulses experienced by the spins. However, the accuracy of the measured excitation profiles is limited by the ESR signal linewidth, and this method cannot provide a well-defined unitary fidelity that describes the gate performance. Moreover, there are other important sources of error beyond the imperfection of the pulse excitation profile. On the other hand, RB provides a quantitative characterization of gates in terms of the average fidelity of the unitary operation, which is a relevant quantity in the context of fault-tolerant QIP. Furthermore, it is unclear how to quantify multi-qubit operations in both the CP/CPMG method in refs. \cite{CPMGPRA, CPMGPRL} and the transfer function method in ref. \cite{AWG1}. Therefore, RB is a more general characterization protocol that produces a quantity that can be directly compared to fault tolerance thresholds and to fidelities measured in other QIP implementations. RB has been applied in a wide range of QIP implementations, including trapped ions~\cite{PhysRevA.77.012307,RBM_Biercuk2009,PhysRevA.84.030303}, liquid state NMR~\cite{colmbm}, superconducting qubits~\cite{RBM_Schoelkopf,RBM_Martinis}, atoms in optical lattices~\cite{RBM_opticallattice}, $^{31}$P donor in silicon~\cite{silicon1,silicon2,RBM_silicon}, and nitrogen-vacancy
(NV) centres in diamond \cite{nv1,nv}. However, as yet RB results in a conventional pulsed ESR system have not been reported. 

In this work, we implement a single qubit RB protocol using pulsed ESR at X-band on a sample of gamma-irradiated fused quartz. The stable, spin-1/2 defect is an unpaired electron at an oxygen vacancy. In order to have the flexibility and precision of control necessary for QIP with arbitrarily shaped optimal control pulses, the spectrometer was custom built and includes an AWG and I-Q modulator for pulse generation, and a specially designed loop-gap resonator (LGR) for efficient, broadband control that accomodates small samples. One challenge with pulsed ESR that is not usually encountered in NMR is that the bandwidth of the microwave resonator that interacts with the sample is comparable to (or narrower) than the desired control bandwidth. Moreover, impedance matching of all elements in the pulse generation, amplification and transmission train is more challenging at microwave frequencies. Here, we identify and partially correct sources of gate error due to pulse distortions; these distortions arise both due to the finite resonator bandwidth and to hardware imperfections. Additionally, we narrow the distribution of local fields that lead to inhomogeneous dephasing ($T_2^*$) by applying a selection sequence, and this demonstrates the dominant role of the $T_2^*$ process as a source of incoherent gate error. The initial RB result of $>6\%$ error probability per gate is significantly reduced to about $0.8\%$ after identifying and mitigating several sources of error, with the lowest error rate being dominated by $T_2^*$ dephasing. Our work provides a first benchmarking of experimental unitary gate fidelities in a conventional pulsed ESR system. This result can be compared, in an unbiased way, with the gate fidelities obtained in other QIP implementations. The use of RB in concert with simulations and the selection sequence sheds light on the contributions of different error sources - $T_1$, $T_2$ and $T_2^*$ processes, $\bm{B}_1$ field inhomogeneity, as well as unitary errors. This is a powerful diagnostic tool that should be broadly useful in ESR applications that increasingly rely on sophisticated pulse sequences that demand high precision; such sequences will benefit from pulses with high unitary fidelities, i.e. pulses that work as desired, independent of the input state. 

\section{Instrumentation}
\label{sec:inst}
\subsection{X-band Pulsed ESR Spectrometer}

\begin{figure}[h!]
\centering
\includegraphics[width=0.45\textwidth]{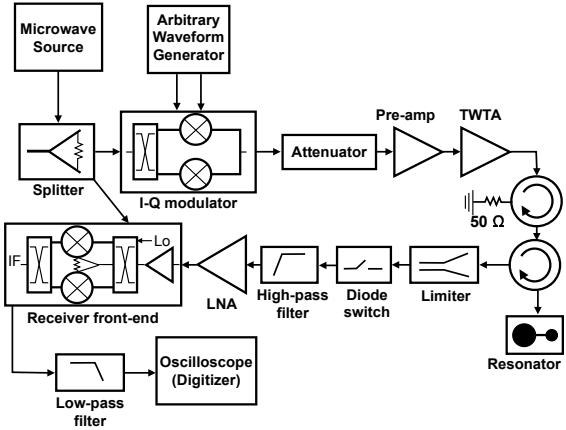}
\caption[Home-built X-band ESR spectrometer design]{\label{fig:ESRcircuit}Schematic of the home-built X-band ESR spectrometer. A signal generated from the microwave source is mixed at the I-Q modulator with $0^{\circ}$ and $90^{\circ}$ phase-shifted components of a shaped pulse from the arbitrary waveform generator. The I-Q modulator outputs the shaped pulse at upconverted frequency, and the pulse phase is accurately controlled by a method explained in the text. A TWT amplifier combined with a pre-amplifier and an attenuator provides an output power up to 500 W. The amplified pulse is transmitted to the loop-gap resonator containing the sample. In the receiver, the ESR signal is mixed with the reference frequency and downconverted to an IF signal, which is digitized by a fast oscilloscope.}
\end{figure}

A schematic of the custom X-band pulsed ESR spectrometer is depicted in Fig.~\ref{fig:ESRcircuit}. The microwave source (Rohde and Schwarz SMF100A) provides a continuous wave (CW) output at $\sim \omega_0/2\pi\sim10$ GHz, and features low phase noise and output power up to $+25$ dBm. In order to generate arbitrary shapes such as GRadient Ascent Pulse Engineering (GRAPE) pulses~\cite{khaneja2005optimal}, we use a single-sideband (SSB) upconversion technique~\cite{zhang2010universal,chamilliard2011electron} with an I-Q modulator (Marki IQ-0714LXP) as the SSB mixer. A $1.2$ GS/s arbitrary waveform generator (AWG, Tektronix AWG5014B with a memory expansion to 32 Mpts) provides both $0^{\circ}$ and $90^{\circ}$ phase-shifted pulse inputs from two output channels, with intermediate frequency (IF)  $\omega_{IF}/2\pi$ ranging from 150-400 MHz. Applying IF signals of the same amplitude but phase-shifted by $\pi/2$ at the I and Q ports suppresses the phase error of the output pulse due to the non-linear power responses of the I and Q mixers. The phase of the output microwave pulse is then controlled by the IF phase generated by the AWG, accurate up to 1 part in 16384.

The I-Q modulator outputs only the upconverted signal at $\omega_0+\omega_{IF}$, and the lower sideband is suppressed. The signal is amplified by a travelling wave tube (TWT, Applied Systems Engineering Model 117) amplifier before being transmitted to the resonant sample cavity. Although the maximum output power of the TWT is $\sim$1 kW, we typically only require tens of Watts in these experiments. Prior to the TWT, a variable attenuator and a low-gain solid-state amplifier (Miteq AFS3-08001200-10-10P-4) are used in order to adjust the input power level. The output of the TWT is directed by a circulator and travels to the loop-gap resonator (see Sec.~\ref{sec:LGR}) that contains the sample, where the pulse produces an oscillating magnetic field $\bm{B}_1$. The circulator has a peak power limit of 500 W, hence the TWT input is adjusted so that its output will not exceed this value. The ESR signal is directed by the circulator to a low-noise preamplifier (Miteq AMF-5F-08001200-09-10P) and then to the receiver (Miteq ARM0812LC2C), where it is mixed with the reference frequency $\omega_0$ and downconverted to $\omega_{IF}$. The receiver includes a second stage of low-noise amplification. A diode switch (Advanced Technical Materials S1517D) is used to protect the receiver from damage during pulsing, and is controlled by a marker channel of the AWG. A diode limiter protects the switch, which is also easily damaged at high microwave power. After the switch, a high-pass filter removes switching transients. Finally, the ESR signal is digitized using a fast oscilloscope (LeCroy WavePro 715Zi) and is recorded on a computer for further processing.

\subsection{Loop-gap Resonator}
\label{sec:LGR}

\begin{figure}[h!]
\centering
\includegraphics[width=0.4\textwidth]{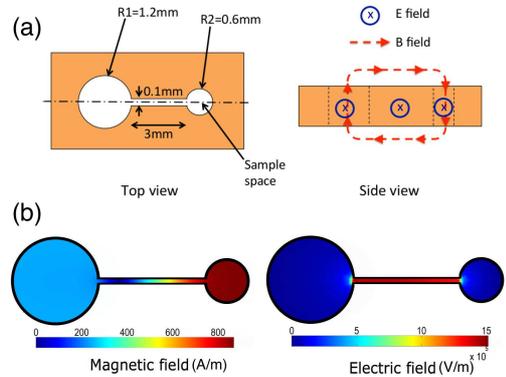}
\caption[Schematic of loop-gap resonator]{\label{fig:LGR} (a) Schematic of the loop-gap resonator. It has two loops and one gap, and is made of copper. The geometry is designed to give a resonance frequency close to 10 GHz, and a quality factor $Q\sim 250$. The magnetic and electric field directions are shown in the side view. The sample is placed inside the smaller loop. (b) High Frequency Structural Simulator (HFSS) simulations of field distributions at the resonance frequency, with input power 1 W. The electric field is the strongest in the gap while the magnetic field is the strongest in the loop. } 
\end{figure}

The design of our loop-gap resonator (LGR) was adapted from Ref.~\cite{Qband_LGR}, with a resonance frequency close to 10 GHz. It is a lumped-element resonator with inductance and capacitance determined by its geometry. The advantages of the LGR compared to conventional cavity resonators is a larger filling factor, large $\Vert\bm{B}_1\Vert$ per square root power, good $\bm{B}_1$ uniformity over the sample volume and larger control bandwidth (with shorter pulse ringdown times) due to a lower quality factor $Q$ ~\cite{chamilliard2011electron,LGR_1}. Figure~\ref{fig:LGR} shows a schematic of the two-loop, one-gap resonator employed here. The loops and the gap are cut by wire electric discharge machining from a 9.6 mm $\times$ 5.4 mm $\times$ 2.5 mm oxygen-free copper block. The loop radii are 1.2 mm and 0.6 mm, and the gap is 0.1 mm wide and 3 mm long. The geometry of the LGR was chosen for a 10 GHz resonance frequency and $Q\sim$250. The microwave electric field is the strongest in the gap while the magnetic field is strongest in the smaller (sample) loop.

\subsection{Probehead}
\label{sec:probe}
The resonator is placed in a rectangular copper enclosure of dimensions 9.4 mm $\times$ 9.6 mm $\times$ 5.4 mm. The lowest resonance frequency of the box is well above 12 GHz, so it does not interfere with the LGR resonance. The inner conductor of a coaxial cable is formed into a one-turn loop and brought close to the larger of the two LGR loops, where it inductively couples the microwave signal to the LGR.  The outer conductor of the coaxial cable is copper, and the inner conductor is silver-plated copper. Impedance matching is done by fine adjustment of the position of the LGR relative to the coupling inductor. 

\begin{figure}[h!]
\centering
\includegraphics[width=0.40\textwidth]{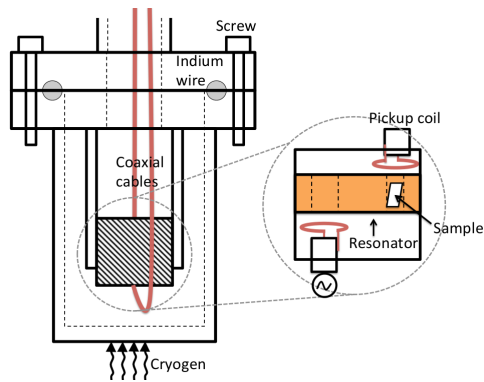}
\caption[Home-built low-T probe]{\label{fig:probe} Schematic of the custom-built ESR probe. The two-loop, one-gap resonator is placed in a rectangular copper box, and the box is placed in a cylindrical copper enclosure which makes contact with cold helium vapour when operated cryogenically. The cylindrical enclosure is sealed using indium wire and screws, and is evacuated. The microwave transmission line is coupled to the LGR by an inductive loop. Tuning and matching of the LGR resonance are accomplished by fine adjustment of the LGR position relative to the coupling inductor. A second coaxial cable is terminated with a pickup coil to detect the microwave field in the vicinity of the sample loop during a pulse.}
\end{figure}

The $\bm{B}_1(t)$ control field seen by the sample does not exactly match the desired pulse shape due to non-linearities in the pulse generation and amplification, and due to the finite bandwidth ($\sim\omega_0/\left(2\pi Q\right)$) of the resonator. In order to identify and correct these errors, we insert a second, smaller coaxial cable terminated with an inductive loop as a pickup coil, as shown in Fig.~\ref{fig:probe}. This pickup coil measures the microwave field in the vicinity of the sample.  The signal from the pickup coil is demodulated in the same way as an ESR signal, by mixing with the reference frequency. The pulse measured by the pickup coil is compared with the desired pulse, and the input waveform adjusted to minimize the difference between the two. This reduces pulse error substantially; however, we note that this procedure assumes the transfer function of the pickup coil is flat over the control bandwidth, and that the pickup coil only couples to the resonant mode and does not itself introduce new modes. We take these assumptions to be approximately true in practice (violations are probably small enough to be ignored in this work, but can be explored in future work). 

The rectangular enclosure containing the resonator is held inside a cylindrical probe-head made of copper, with 28 mm inner diameter and 85 mm inner depth. A schematic of the probe is illustrated in Fig.~\ref{fig:probe}. It is designed to fit into the Oxford Instruments CF935O continuous-flow cryostat, and can be cooled to slightly below 4 K by pumping on the Helium space. Experiments at cryogenic temperatures have been carried out with this setup, and electron spin polarizations following the expected Curie law were observed down to about 5 K.  However, the experiments reported in this paper were carried out at room temperature, since our goal here is to characterize the precision of spin manipulation by the microwave control field. In future work we plan to perform RB experiments at low temperature and on multi-qubit samples. 

\section{Single Qubit Randomized Benchmarking}
\label{sec:sgp}
The strategy of RB is to average over a set of random sequences of computational gates, with each sequence followed by a reverse (recovery) operation that makes the entire evolution the identity operator, in the absence of gate error. The computational gate set should be chosen such that it forms a depolarizing channel upon averaging, which is true as long as the errors are mostly gate-independent~\cite{emersonRBM,PhysRevLett.106.180504,PhysRevA.85.042311}. The depolarizing parameter $p$ of the averaged channel $\bar{\Lambda}$ is given by:
\begin{equation}
\bar{\Lambda}=(1-p)\rho+\frac{p}{D}\one
\end{equation}
and is related to the average gate fidelity. Here $D$ is the dimension of $\rho$. Averaging over the full unitary group provides a benchmark for control fidelity over the complete set of quantum gates. However, since the unitary group is a continuous set with the number of parameters increasing exponentially in the number of qubits $n$, generating an arbitrary unitary operator is exponentially hard with increasing $n$. Therefore, it is more practical to use a discrete set of gates. It is desirable that the operations can be efficiently (classically) simulated and can be generated from a small set of elementary one and (in the general case) two-qubit gates. The Clifford group, denoted as $\mathcal{C}$, is a set that is attractive for RB for several reasons. First, a universal gate set can be generated using $\mathcal{C}$ with the addition of a so-called magic state together with measurement in the computational basis, which means that there exists a universal quantum computation model in which all necessary gates can be drawn from the Clifford group~\cite{magicstate}. Furthermore, benchmarking the Clifford gates is useful since most encoding schemes for fault tolerant QIP are based on stabilizer codes, in which the error correction is performed using Clifford gates.

We followed the single-qubit RB protocol implemented in a liquid-state NMR experiment presented in Ref.~\cite{colmbm}. The single qubit Clifford randomization is equivalent to a randomization using the 48 operations parametrized as~\cite{colmbm,MartinThesis,PhysRevA.79.042328}
\begin{align}
\label{eq:CliffOp}
\mathcal{S}\mathcal{P}&=\exp\left(\pm i\frac{\pi}{4}Q\right)\exp\left(\pm i\frac{\pi}{2}V\right),\\\nonumber Q&\in \lbrace\sigma_x,\sigma_y,\sigma_z\rbrace, V\in \lbrace\mathbf{1},\sigma_x,\sigma_y,\sigma_z\rbrace.
\end{align}
In this protocol, the $\pi/2$ rotations (which belong to the symplectic group and are denoted as $\mathcal{S}$ in Eq.~\ref{eq:CliffOp}) are computational operations while the Pauli operations (which belong to the Pauli group and are denoted as $\mathcal{P}$ in Eq.~\ref{eq:CliffOp}) serve to toggle the Pauli frame and thus depolarize the noise. $\mathcal{S}\mathcal{P}$ realizes an operation in the Clifford group. To quantify the average error per gate, randomly chosen sequences of $L$ Clifford operations in Eq. (\ref{eq:CliffOp}) are applied to a fixed, known initial state. A natural choice for the initial state is the thermal state, which can be simply written as $\rho_i=\sigma_z$ since the identity part of the density matrix is irrelevant in both unitary evolution and in measurement. 

\begin{figure}[h!]
\centering
\includegraphics[width=0.45\textwidth]{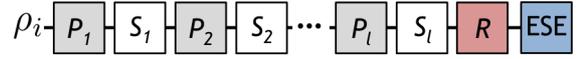}
\caption[Randomized benchmarking pulse sequence]{\label{fig:BMcircuit} A particular realization of a Clifford gate sequence in the RB protocol. $P$ and $S$ indicate Pauli and computational operations, respectively, and $R$ is the recovery gate that brings the state to either $+\sigma_z$ or $-\sigma_z$. The sequence is truncated at some length $l$, and the final state is measured by the amplitude of the electron spin echo after a read-out pulse sequence, as described in the text. The electron spin echo measurement part is denoted by ESE in the figure.}
\end{figure}

The initial state is tracked through a sequence of Clifford operations (Eq.~\ref{eq:CliffOp}), and the sequence is truncated at different lengths $l\le L$ to measure the fidelity decay curve. At each truncation $l$, a recovery gate is chosen at random to return the state to either $\pm \sigma_z$. Then, a read-out pulse sequence that contains a $\pi/2$ rotation pulse, a 700 ns delay and a $\pi$ rotation pulse, is used to generate a spin echo. The amplitude of the spin echo in the time domain, equivalent to the integrated intensity of the corresponding absorption peak in the frequency domain, is the quantity we measure. In the experiment, we pre-calculate the final state prior to the spin echo measurement, and choose the phase of the $\pi/2$ read-out pulse so that the positive eigenvalue is always measured. The quantum circuit implementing a particular series of Clifford operations is shown in Fig.~\ref{fig:BMcircuit}. 

The single qubit RB protocol can be summarized as follows~\cite{colmbm}:
\begin{enumerate}
\item Choose a maximum number of Clifford operations $L$, and a set of random integers $l=\lbrace l_1<\ldots<L\rbrace$ that is the length of a truncated pulse sequence (i.e. a subsequence). The number of elements in $l$ is denoted $N_l$.

\item Generate $N_g$ random sets of $L$ computational gates, and truncate each sequence at length $l_k\in l$.

\item For each subsequence of length $l_k$ (total $N_g\times N_l$ subsequences), do the following:
\begin{enumerate}
\item Generate $N_p$ random sets of Pauli gates of length $l_k+2$.
\item Interleave computational gates with the $l_k+1$ Pauli gates.
\item Calculate recovery operation and insert it after $(l_k+1)^{th}$ Pauli gate.
\item Insert final Pauli randomization gate before the spin echo detection.
\item Calculate the phase of the spin echo read-out pulse that yields the positive eigenvalue of $\sigma_z$.
\end{enumerate}
\item Evolve the known initial state under each of the $N_g\times N_l\times N_p$ number of random pulse subsequences.
\item Measure the electron spin echo amplitude at the end of each subsequence and compare with the same measurement on a reference state.
\item For each $l_k$ where $k=1\ldots N_l$, calculate the average remaining spin magnetization along $\sigma_z$ by averaging over $N_g\times N_p$ subsequences.
\item Plot the average remaining signal as a function of $l$, and fit the curve to $f_l=\alpha (1-p)^l$, where the average gate error is $p/2$. The state preparation and measurement (SPAM) error is absorbed in the constant $\alpha$.
\end{enumerate}

As indicated in steps 6 and 7, the average gate error is calculated from the decay rate of $\langle\sigma_z\rangle$ as a function of the number of gates $l$.

\section{Simulations and Initial Experiment}
\begin{figure}[h!]
\centering
\subfloat[\label{fig:t2star}Thermal-state ESR spectrum]{
\epsfig{file=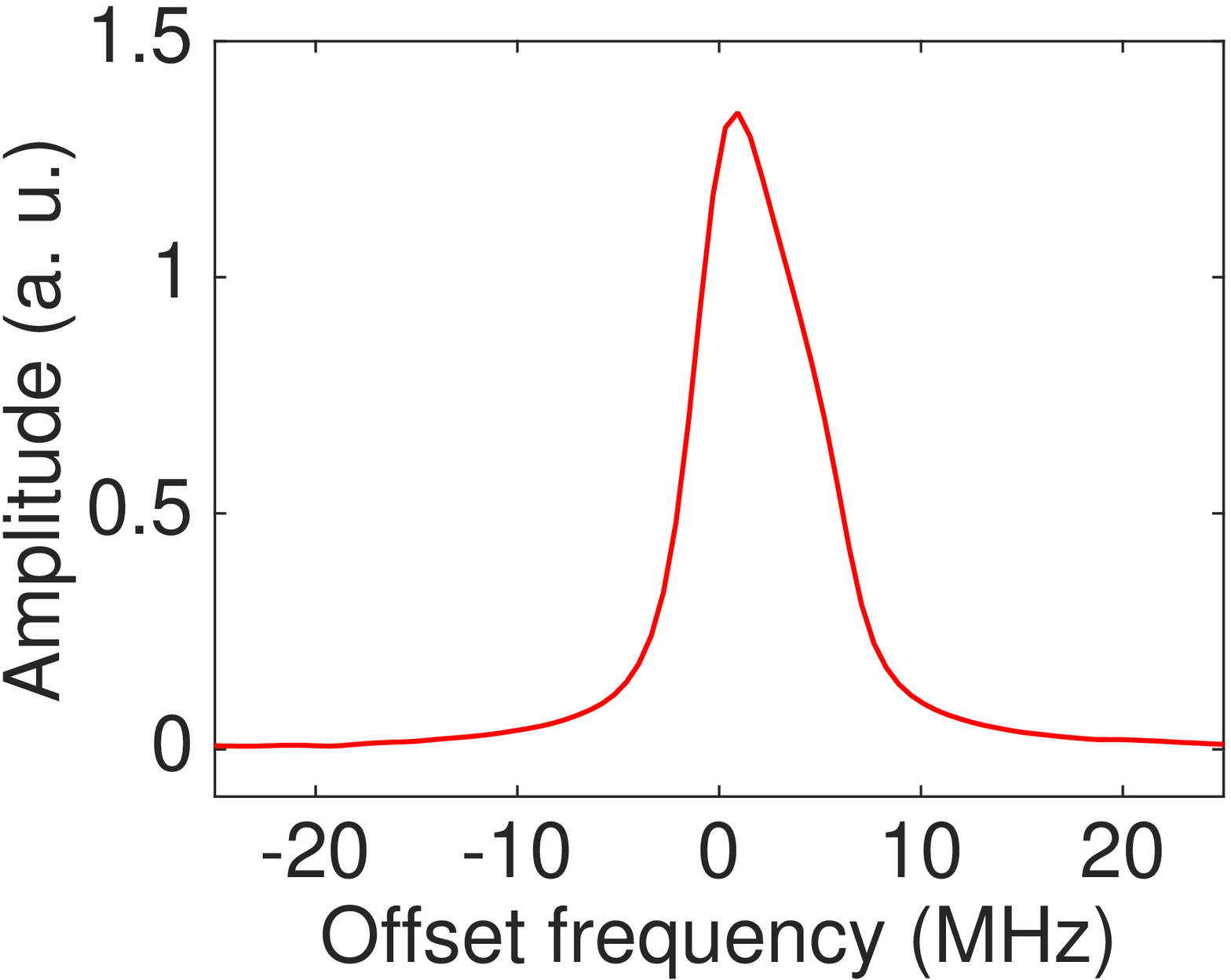,angle=0,width=0.24\textwidth}
}
\subfloat[\label{fig:rfdist}$\bm{B}_1$ distribution]{\hspace{-2mm}
\epsfig{file=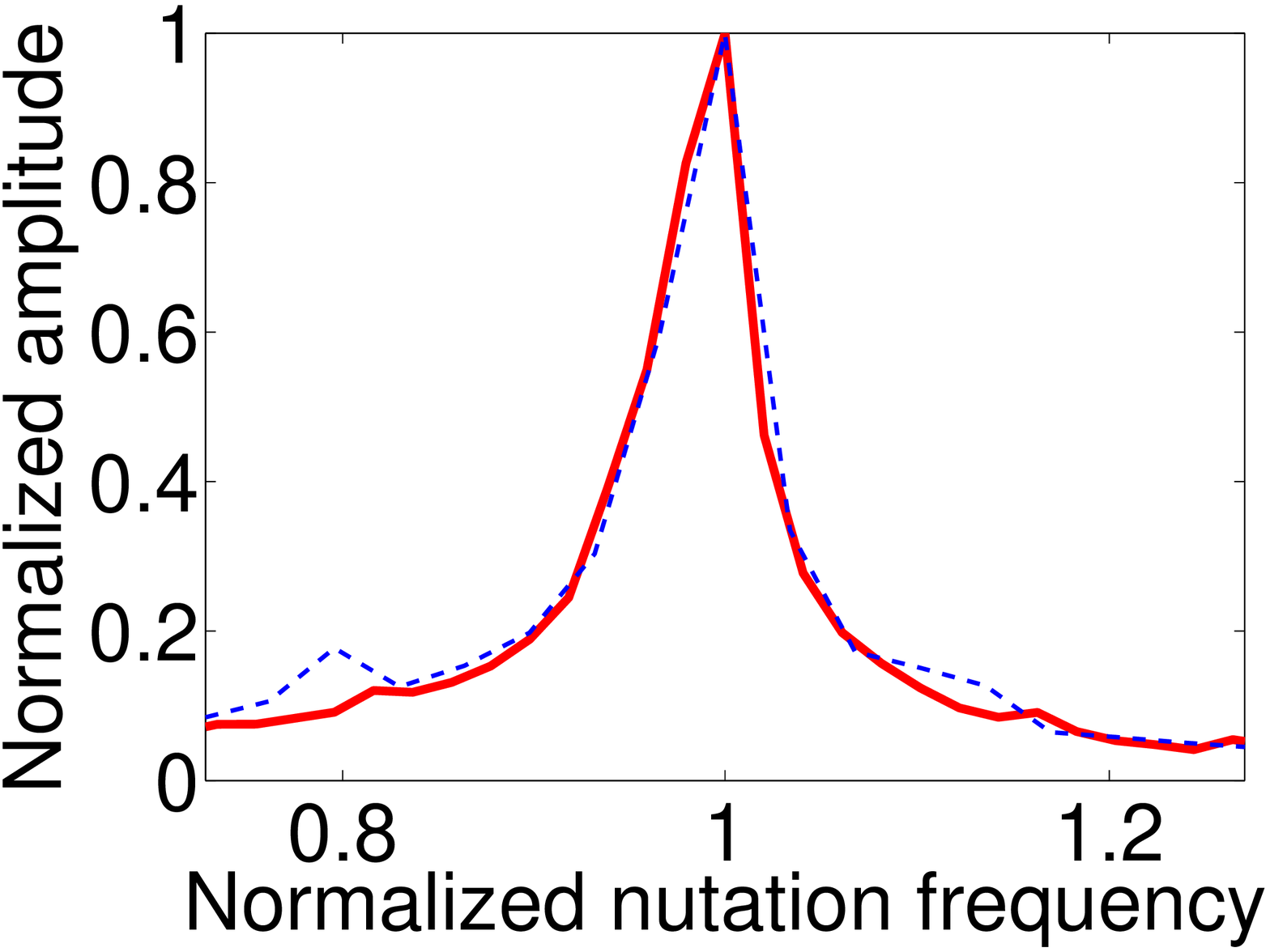,angle=0,width=0.24\textwidth}
}
\caption[$T_2^{*}$ and $\bm{B}_1$ field distribution of the irradiated fused quartz measured in the home-built ESR spectrometer]{\label{fig:dist}(a) Room temperature ESR spectrum of irradiated fused quartz obtained by Fourier transforming the spin echo signal.  The linewidth is mainly determined by the anisotropy of the g-value and inhomogeneity of the static magnetic field. The asymmetric line shape indicates the powder pattern of the quartz sample. (b) $\bm{B}_1$ distribution profile obtained by taking the Fourier transform of the experimental (red solid line) and simulated (blue dashed line) Rabi oscillations, as described in the text. The nutation frequency is normalized by the peak frequency of the distribution, $31.7$ MHz. The FWHM is about 2.1 MHz.}
\end{figure}

\begin{figure}[h!]
\centering
\includegraphics[width=0.475\textwidth]{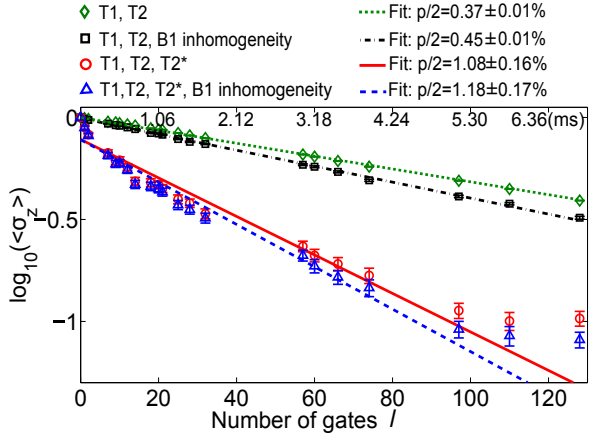}
\caption[]{\label{fig:bmsim} Simulated RB results on the single qubit (fused quartz) spin system, plotted in semi-log scale. The lower x-axis is the number of gates and the upper x-axis gives the actual sequence time. The expectation value $\langle \sigma_z\rangle$ is measured after application of $l$ randomized gates and fitted using $f_l=\alpha (1-p)^l$. The expectation value $\langle \sigma_z\rangle$ for each value of $l$ is the result averaged over $N_g\times N_p=98$ random sequences. The simulations use experimentally measured values of $T_1$, $T_2$, and $T_2^{*}$ at room-temperature, which are 160 $\mu$s, 5 $\mu$s, and 60 ns, respectively. All curves include the $T_1$ and $T_2$ processes of the electron, through a Lindblad formalism. The $T_2^{*}$ and $\bm{B}_1$ distributions are included as follows: Green ($\diamond$) indicates no distributions; Black ($\square$) indicates $\bm{B}_1$ only; Red ($\circ$) indicates $T_2^{*}$ only; blue ($\triangle$) indicates both $\bm{B}_1$ and $T_2^{*}$. The size of the error bars represents standard error of the mean averaged over the 98 sequences.}
\end{figure}

The single qubit system employed for the benchmarking experiment is gamma-irradiated fused quartz~\cite{quartz}, a paramagnetic sample in powder form where the primary defect is an unpaired electron at an oxygen vacancy. We obtained $T_1=160$ $\mu$s, $T_2=5$ $\mu$s, and $T_2^*=60$ ns from ESR experiments at room-temperature. $T_2^{*}$ is determined from the linewidth of the thermal-state ESR spectrum in the frequency domain.   The $T_2^{*}$ line-broadening here is mainly due to the anisotropy of the $g$-value, which produces a powder pattern of width close to 5 MHz, and also to inhomogeneity of the applied static magnetic field. Different spins within the sample volume also experience different nutation frequencies due to the inhomogeneity of the applied microwave field ($\bm{B}_1$) in the resonator loop. To obtain the $\bm{B}_1$ distribution profile, we measure the time-domain Rabi oscillations. Simulations show that $T_2^*$ does not contribute significantly to the Rabi decay envelope for this sample, and the decay is dominated by $\bm{B}_1$ inhomogeneity. The insensitivity of the Rabi oscillations to $T_2^{*}$ noise agrees with the intuition that the continuous pulse partially refocuses the dynamics due to static field inhomogeneity. Therefore, the Fourier transform of the time-domain Rabi oscillations gives a good estimate of the $\bm{B}_1$ distribution over the sample. This distribution is confirmed by comparing experimental and simulated Rabi decays, which allows us to refine the estimate for the $\bm{B}_1$ distribution profile. The ESR line-broadening due to $T_2^{*}$ and the $\bm{B}_1$ distribution profile are both shown in Fig.~\ref{fig:dist}. Using the decoherence parameters and the $\bm{B}_1$ inhomogeneity data, we simulated a RB experiment with $N_g=7$, $N_p=14$ giving a total of 98 sequences, and $l=\lbrace$1, 2, 7, 9, 10, 12, 14, 18, 20, 21, 25, 28, 32, 57, 60, 66, 74, 97, 110, 128$\rbrace$ chosen at random. Both computational and Pauli gates are realized by 35 ns long Gaussian microwave pulses, yielding Clifford operations that are 70 ns in total length. Figure ~\ref{fig:bmsim} summarizes the simulation results under four different conditions: (1) no $\bm{B}_1$ inhomogeneity and no $T_2^{*}$ effect, (2) with $\bm{B}_1$ inhomogeneity and no $T_2^{*}$ effect, (3) with $T_2^{*}$ effect and no $\bm{B}_1$ inhomogeneity, and (4) with both $\bm{B}_1$ inhomogeneity and $T_2^{*}$ effects. $T_1$ and $T_2$ processes are included in all four simulations by using a Lindblad model, while the local field inhomogeneity effects are included in (2)$\sim$(4) using the method of weighted averaging over multiple simulations. In the absence of any local field inhomogeneities, the simulated error per gate is $0.37\%$. Since the controls fields are ideal in this case, the imperfection is solely due to $T_1$ and $T_2$ processes. When the local field inhomogeneities are included, the error per gate increases to $0.45\%$ for $\bm{B}_1$, $1.08\%$ for $T_2^*$ and $1.18\%$ for both $\bm{B}_1$ and $T_2^*$. Thus, the simulations clearly show that the $T_2^*$ effect contributes more significantly to the gate error rate than $\bm{B}_1$ inhomogeneity for this sample. This is not surprising, since the $\bm{B}_1$ field distribution is narrower than the static field distribution (see figures ~\ref{fig:t2star} and ~\ref{fig:rfdist}). 

With no optimization of the microwave pulses beyond simple power calibrations, RB was implemented experimentally in the quartz single qubit system. The error per gate was $\sim 6\%$, indicating a relatively poor level of quantum control. In the following section, we discuss identification and mitigation of pulse distortions and other factors that allow the gate error rate to be significantly reduced. 

\section{Pulse Corrections}
\begin{figure}[h!]
\centering
\subfloat[\label{fig:PTC1a}Pulse shape]{
\includegraphics[width=0.24\textwidth]{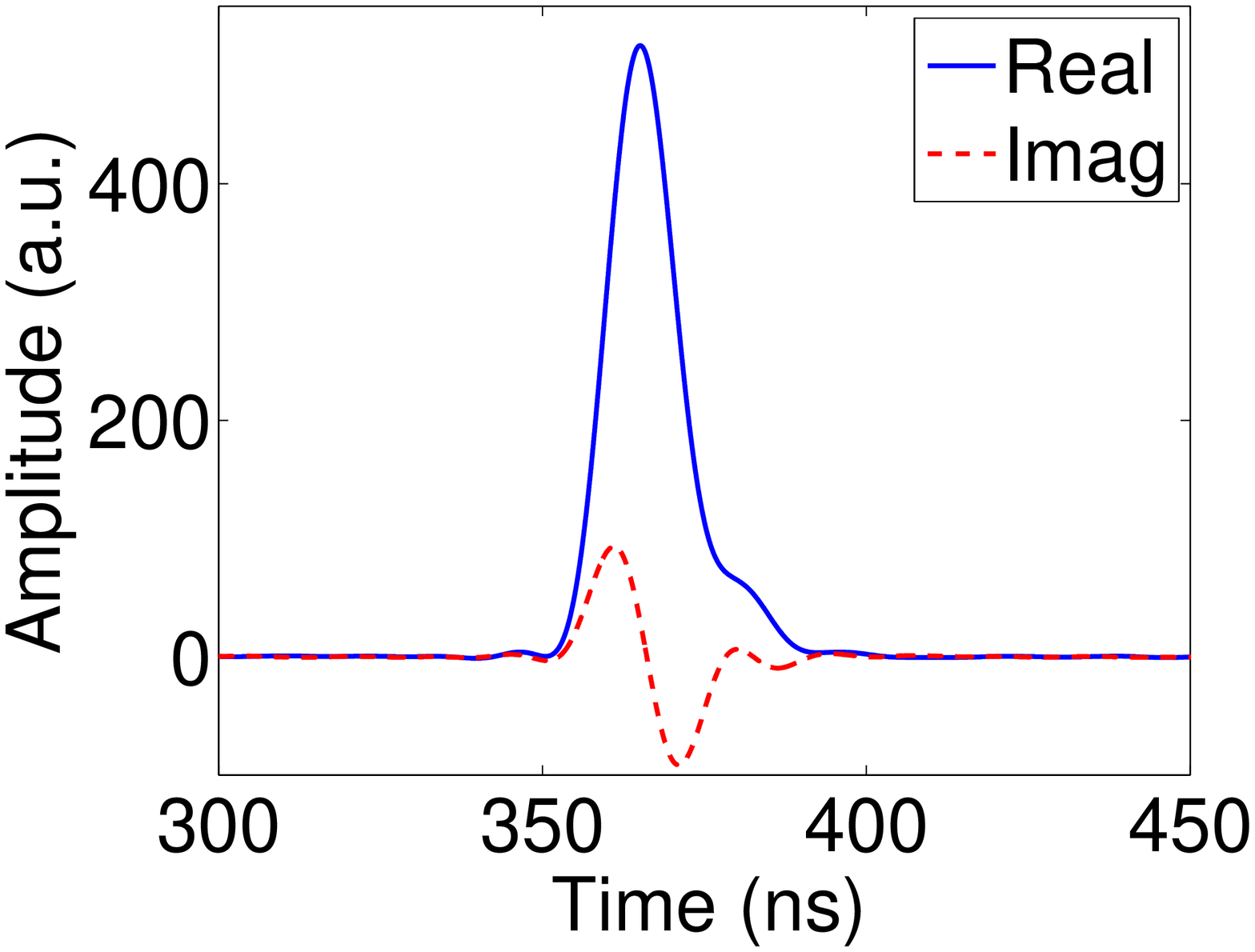}
}
\subfloat[\label{fig:PTC1b}Calibration data]{\hspace{-2mm}
\includegraphics[width=0.24\textwidth]{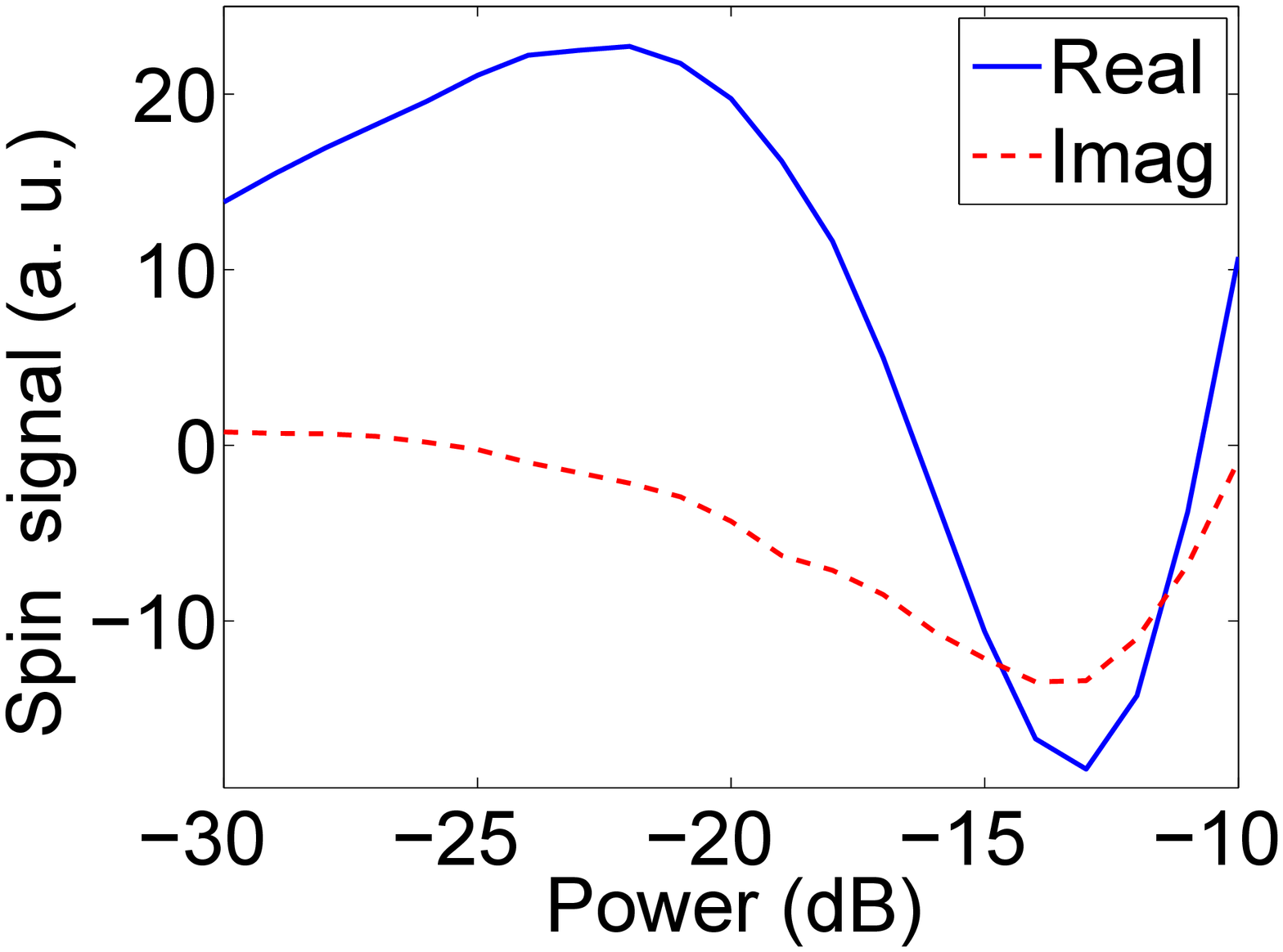}
}
\caption[Phase transient error]{\label{fig:PTC1}(a) The shape of a $35$ ns Gaussian pulse measured using the pickup coil shown in Fig.~\ref{fig:probe}. The pulse is applied along the rotating frame y-axis. The solid curve represents the real part (y-axis component)  of the measured pulse and the dashed curve is the imaginary part (x-axis component). Ideally, the imaginary component of the pulse is zero, however the pickup coil reveals a large phase transient. (b) Spin magnetization signal measured as a function of the pulse power. The Gaussian pulse is applied along the y-axis to the spin state $\sigma_z$, and is followed by a 700 ns delay and a $\pi$ pulse to form an echo. Without pulse imperfections, only the real part (x component, solid curve) of the spin signal should oscillate as a function of the pulse power while the imaginary part (y component, dashed curve) should be zero. However, the phase transient leads to an undesired spin rotation out of the x-z plane.}
\end{figure}

The pickup coil introduced in Sec.~\ref{sec:probe} directly measures the microwave field in the vicinity of the sample. The largest pulse imperfection revealed by the pickup coil was a phase transient error. Here we use a $35$ ns Gaussian pulse as an example. The pulse is applied along the rotating frame y-axis, so that the amplitude of the imaginary component (x-axis component) of the pulse is zero. However, as shown in Fig.~\ref{fig:PTC1a}, the measured output pulse has a noticeable phase transient with nonzero imaginary component. This transient effect leads to undesired spin dynamics as shown in Fig.~\ref{fig:PTC1b}.  

\begin{figure}[h!]
\centering
\subfloat[\label{fig:PTC2a}Pulse shape]{
\includegraphics[width=0.23\textwidth]{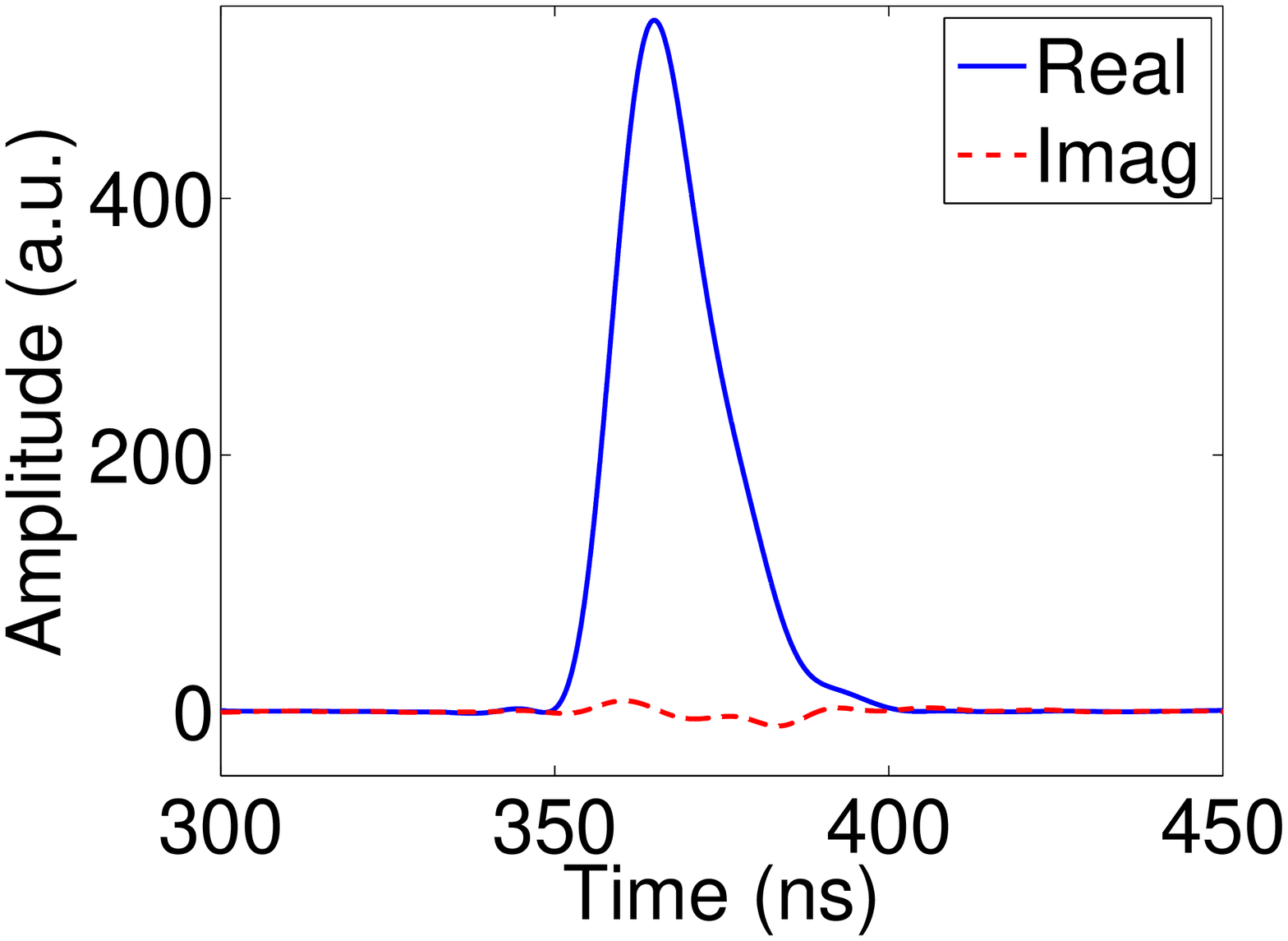}
}
\subfloat[\label{fig:PTC2b}Calibration data]{\hspace{-2mm}
\includegraphics[width=0.23\textwidth]{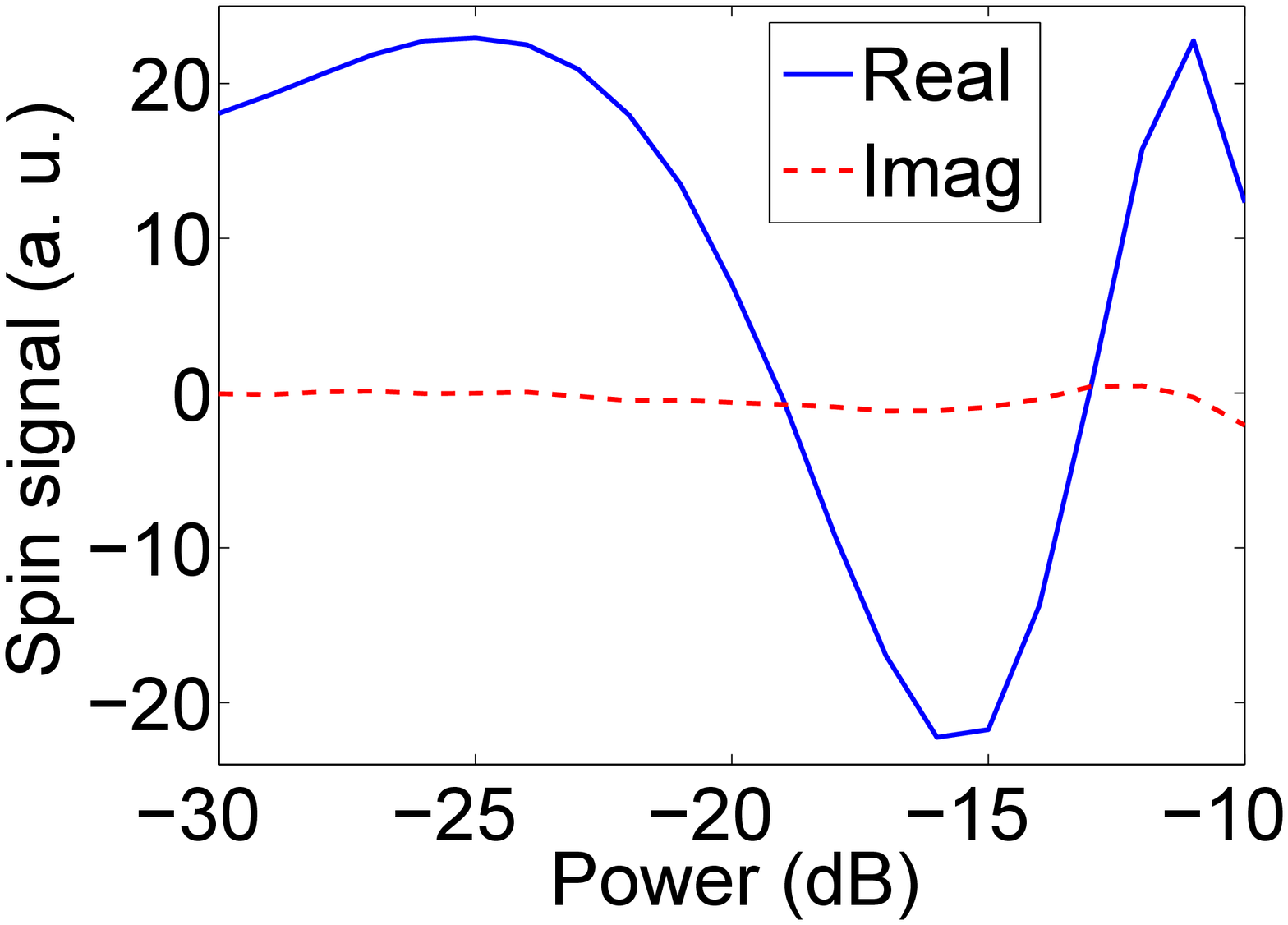}
}
\caption[Phase transient correction]{\label{fig:PTC2}(a) The shape of a $35$ ns Gaussian pulse after implementing the phase transient correction described in the text. The unwanted imaginary component (x-axis component, dashed curve) of the pulse is suppressed.  (b) Spin magnetization signal measured as a function of the power of the PTC pulse.  The deviation of the spin trajectory from the x-z plane (dashed curve) is strongly reduced compared to the uncorrected pulse.}
\end{figure}

Figure~\ref{fig:PTC1b} shows the rotation of the spin magnetization vector as the power of the Gaussian pulse is varied. The y-axis pulse is applied to the spin state $\sigma_z$, and it is followed by a 700 ns delay and a $\pi$ pulse to form a spin echo signal. The solid and dashed curves represent real (x) and imaginary (y) components of the measured spin echo amplitude. In the absence of phase transients, the imaginary part should be zero for all powers as the spin magnetization vector remains within the x-z plane.  However, a nonzero y-component in Fig.~\ref{fig:PTC1b} indicates that the spin is rotated out of the x-z plane by the phase transient. In order to suppress this pulse error, we designed an input pulse with an imaginary part of equal amplitude but opposite sign to cancel the transient. The phase transient corrected (PTC) pulse is designed as follows. First, we take the imaginary part of the pulse measured by the pickup coil, and digitize it so that the time resolution of the measured pulse matches the AWG sampling rate.  We then multiply the imaginary part of the digitized pulse by $-1$ to negate the phase error. However, the digitization step can introduce a scaling error in the pulse amplitude, and thus the multiplication by $-1$ may not exactly cancel the unwanted imaginary component. We find that introducing a scaling factor of $1.07$ to the PTC pulse gives the best result in minimizing the phase transient effect. Figure~\ref{fig:PTC2a} demonstrates that using the PTC pulse, the phase transient is significantly reduced. As shown in Fig.~\ref{fig:PTC2b}, the deviation of the spin evolution trajectory from the x-z plane is strongly suppressed when the same experiment is conducted with the PTC pulse. It should be noted that the real component of the pulse is asymmetric in time, with a slowly decaying tail; this is due to the finite bandwidth of the resonator. However, since the integrated pulse area can be calibrated with high accuracy to yield the desired spin rotation, we only need to correct the distortion in the imaginary part of the pulse. It should also be mentioned that due to the limited resonator bandwidth there is a trade-off between the pulse length and the ability to correct the phase transient. We find that Gaussian pulses of 35 ns length (hence Clifford gates of 70 ns length) offer the best compromise in the particular system reported here. 

\begin{figure}[h!]
\centering
\subfloat[\label{fig:BD300n}$300$ ns TWT unblanking delay]{
\epsfig{file=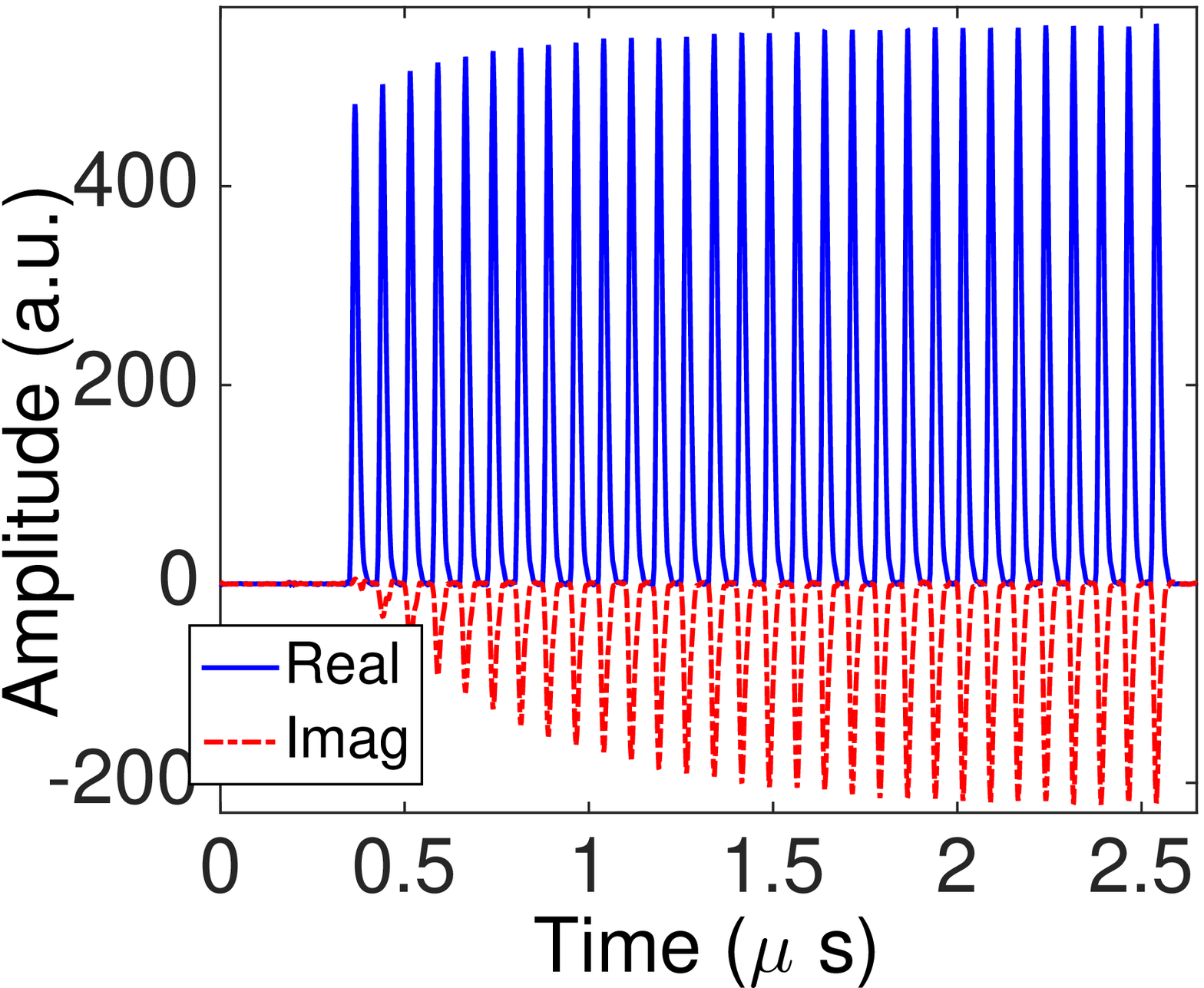,angle=0,width=0.24\textwidth}
}
\subfloat[\label{fig:BD2u}2 $\mu$s TWT unblanking delay]{\hspace{-2mm}
\epsfig{file=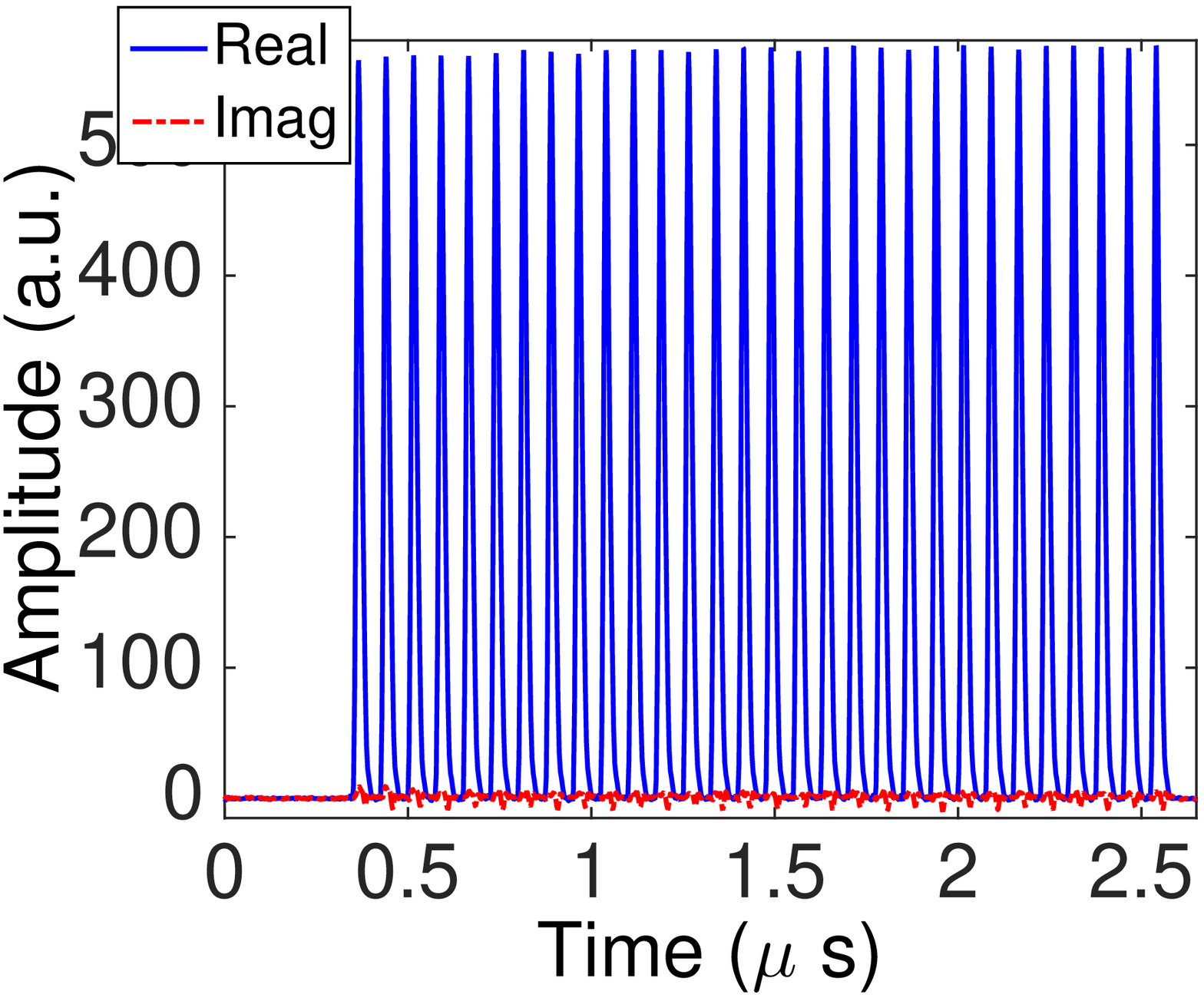,angle=0,width=0.24\textwidth}

}
\caption[Effect of TWT unblanking delay on the pulse shape]{\label{fig:BD} (a) A series of ($35$ ns) PTC Gaussian pulses with 10 ns spacing, measured by the pickup coil, with a $300$ ns TWT unblanking delay. Here $t=0$ corresponds to the unblanking time. The amplitude and phase require about 2 $\mu$s to stabilize after unblanking. (b) The same series of pulses with a 2 $\mu$s unblanking delay, indicating much better stability of both phase and amplitude. Here unblanking occurs at $t=-1.7$ $\mu$s with respect to the time axis shown.  }
\end{figure}

Another significant contribution to the gate error was the phase and amplitude droop of the TWT microwave amplifier. We found this error to be strongly dependent on the amplifier unblanking time delay. Figure~\ref{fig:BD300n} shows a series of  phase transient corrected $35$ ns Gaussian pulses, as measured by the pickup coil, when the TWT unblanking delay is set to $300$ ns. It is clear from the figure that both amplitude and phase are unstable until about 2 $\mu$s after the blanking is turned off (at t=0 in Fig. \ref{fig:BD300n}). Figure~\ref{fig:BD2u} shows that when we use a 2 $\mu$s unblanking time, the amplitude and phase stabilities are much better.

After the phase transient effect is corrected and the TWT unblanking delay is set to 2 $\mu$s, the RB experiment was carried out again, and the error per gate was found to be $1.72\pm0.25\%$ ($\diamond$ symbol in Fig. \ref{fig:bmresults}). This error rate is still about an order of magnitude larger than the rate due solely to $T_1$ and $T_2$ processes, as predicted by simulations. A large part of the remaining error in the experimental value is due to ensemble inhomogeneity effects, which we describe in the next section. 

\section{Local Field Inhomogeneities and Spin Packet Selection}
\label{selection}
The simulation results shown in Fig.~\ref{fig:bmsim} indicate that the average gate fidelity is limited by local field inhomogeneities: the static distribution of Larmor frequencies ($T_2^*$) and the inhomogeneity of the microwave field $\bm{B}_1$ across the sample volume. The $\bm{B}_1$ distribution is relatively narrow, and we observed from simulations that the $T_2^*$ effect contributes more strongly to the error rate. To test this experimentally, we designed a pulse sequence to effectively narrow the static field distribution.

One approach to overcome field inhomogeneity is to use composite pulses~\cite{LevittComposite}. However, useful composite pulses are typically much longer than simple pulses. Given the timescales of $T_2^*$ and $T_2$ in our system, composite pulses do not improve gate fidelities, but worsen them.  For example, at least three simple pulses are required for a spin inversion ($\pi$-rotation) composite pulse that is robust to the $T_2^{*}$ field inhomogeneity~\cite{tycko}, meaning that the total pulse duration of either a Pauli or a computational gate is $3\times35$ ns=$115$ ns. A simulation using the $115$ ns composite pulses shows that even in the absence of the field inhomogeneities, where the error solely comes from $T_1$ and $T_2$ processes, the error per gate in the RB experiment would be $1.08\pm 0.01\%$. Thus, composite pulses were not used in our experiments. 

\begin{figure}[h!]
\centering
\epsfig{file=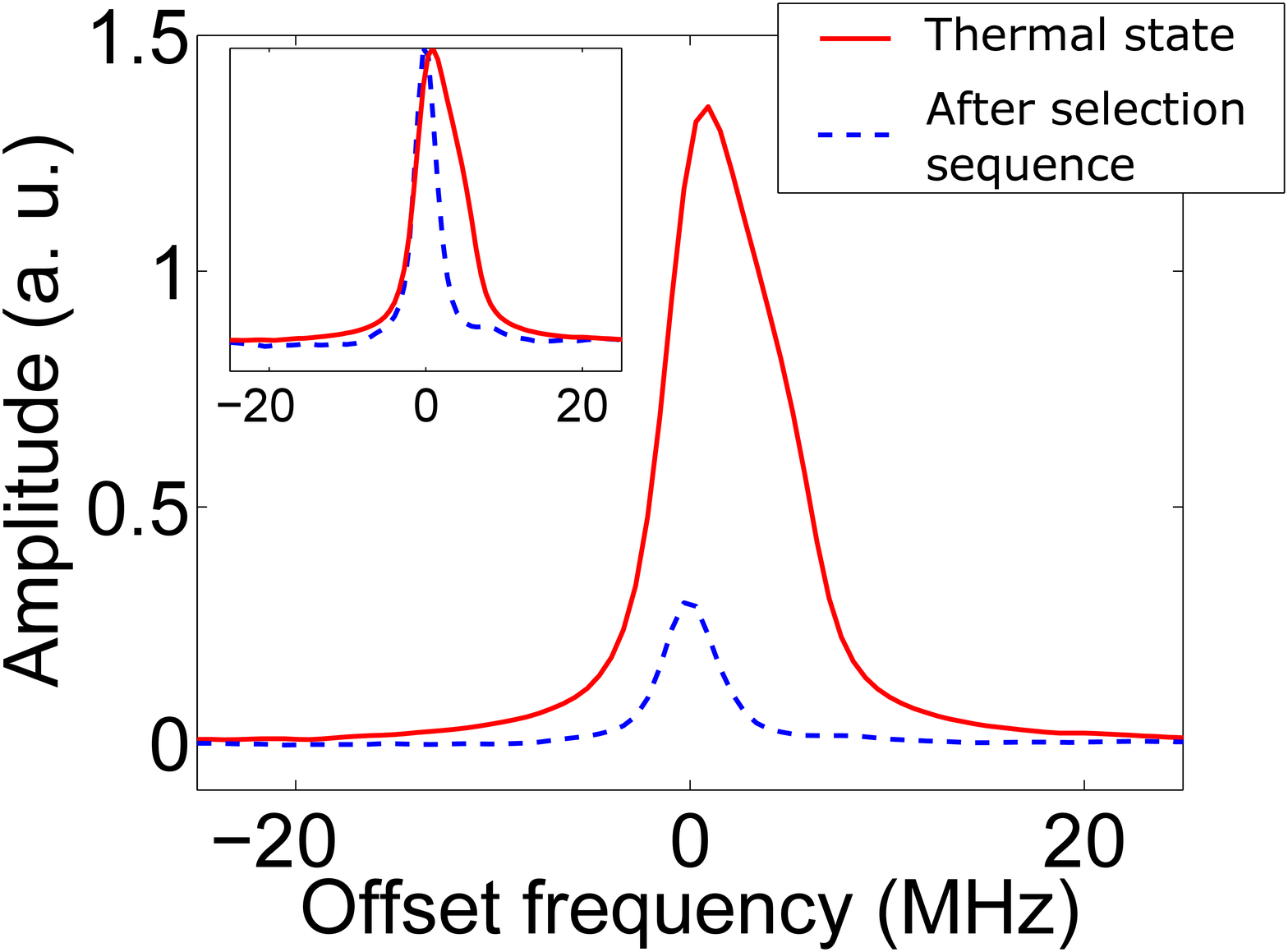,angle=0,width=0.45\textwidth}
\caption[$T_2^{*}$ and $\bm{B}_1$ field distribution of the irradiated fused quartz measured in the home-built ESR spectrometer]{\label{fig:distsel} Comparison of the ESR spectra before (solid) and after (dashed) applying the selection sequence described in the text. The signal intensity is reduced since only a small portion of the spin magnetization remains after the sequence. However, the linewidth is significantly reduced, as shown in the inset where we plot the two spectra normalized to the same amplitude. The linewidths of the two spectra are 5.3 MHz and 2.6 MHz. The sequence dephases spin magnetization at larger frequency offsets while preserving the magnetization of spin packets near zero offset. Hence, the center of the post-selection spectrum shifts towards zero offset.
}
\end{figure}

The approach taken here is to select a subset of the spin ensemble within a narrower static field distribution, similar to the idea of RF selection that has been used successfully in NMR QIP ~\cite{colmbm,Cory199323,Shaka1985175}. The selection sequence is constructed as follows. First, a $400$ ns GRAPE \cite{khaneja2005optimal} pulse (with 400 time points) rotates spins that experience the on-resonance field by $2\pi$ around the x-axis, with a simulated unitary fidelity (Hilbert-Schmidt (HS) norm) of $99.9\%$. For the off-resonance spin packets, the unitary fidelity quickly decreases with offset. Thus, mostly on-resonance spins are left pointing along the z-axis after the GRAPE pulse. 

Next, we wait for a duration $\sim T_2$ in order to dephase the transverse component of the off-resonance spin packets pointing along other directions in the Bloch sphere. By repeating these two steps, but keeping the total sequence time short compared to $T_1$, an initial state is prepared that corresponds to a narrower ESR linewidth. 

Ideally, the dephased off-resonance spin packets will be in a fully unpolarized state and not contribute to subsequent spin signals.  We find that repeating the sequence four times and using a delay equal to $T_2$ provides the best narrowing of the ESR linewidth while retaining a signal level well above the detection noise floor. 

In this post-selection state, $T_2^{*}$ is extended to more than twice its original length, as shown in Fig.~\ref{fig:distsel}. Note that the selection sequence does not necessarily improve the $\bm{B}_1$ field homogeneity. In order to reduce $\bm{B}_1$ field inhomogeneity, spin packets from localized region of the sample need to be selected. We do not see any significant improvement in the $\bm{B}_1$ distribution after this sequence. Nevertheless, we show in the next section that this selection sequence does indeed improve the average gate fidelity, suggesting that the $T_2^*$ process is a dominant source of gate error.

\section{Results and Discussion}

\begin{figure}[h!]
\centering
\includegraphics[width=0.475\textwidth]{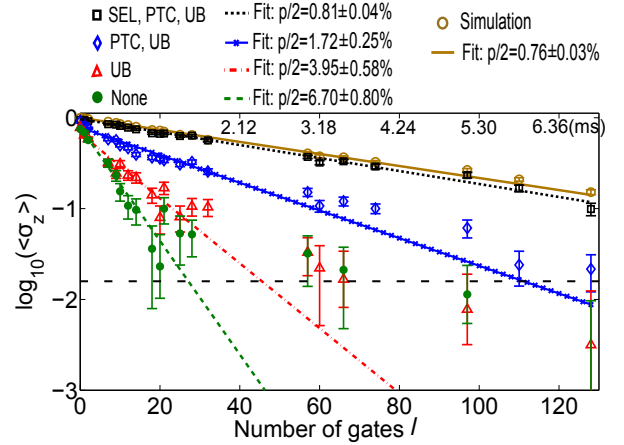}
\caption[Summary of randomized benchmarking experimental results]{\label{fig:bmresults} Summary of experimental results plotted in semi-log scale. The lower x-axis is the number of gates and the upper x-axis gives the actual sequence time. The expectation value $\langle \sigma_z\rangle$ is measured after $l$ randomized gates and fitted using $f_l=\alpha (1-p)^l$. The expectation value $\langle \sigma_z\rangle$ for each value of $l$ is the result averaged over $N_g\times N_p=98$ random sequences. The initial experiment, before making the improvements described in the text, is denoted as `none' in the legend. The average error rate is above $6\%$ ($\bullet$). The triangles ($\triangle$) indicate the result  after setting the TWT unblanking delay to $2$ $\mu$s (denoted UB), where the error probability is reduced to $3.95\%$. Adding phase transient correction (denoted PTC), the error rate is further reduced to $1.72\%$ ($\diamond$). Finally, adding the selection sequence (denoted SEL) to narrow the static field inhomogeneity, the gate error decreases to $0.81\%$ ($\square$). The open circles are the results of the RB simulation. The simulation includes the selection sequence, the $T_1$, $T_2$, $T_2^{*}$ values measured in experiment, and the $\bm{B}_1$ inhomogeneity distribution (but otherwise assumes perfect pulses). The simulation should be compared to the final experimental data with $0.81\%$ error rate. Error bars represent the standard error of the mean averaged over the 98 sequences. The horizontal dashed line (black) corresponds to the experimental noise floor.}
\end{figure} 

The RB experimental results under several different conditions are plotted in Fig.~\ref{fig:bmresults}. The semi-log plot shows the expectation value of $\sigma_z$ measured after applying $l$ Clifford gates. The depolarizing constant as well as the error per gate can be calculated from the decay rate of the expectation value as a function of the number of Clifford operations. 

The initial experimental RB data, without any improvements, is plotted ($\bullet$ symbol) and corresponds to an error probability of 6.7$\%$. Changing the TWT unblanking delay ($\triangle$ symbol) results in an error probability close to 4$\%$. Both of these give decay curves that deviate substantially from a single-exponential fit after about 30 gates. Hence, we only use the first few points to fit to a single exponential and extract the error rate. Further control improvement is achieved by employing phase transient corrected pulses ($\diamond$ symbol). Combining the longer unblanking delay with phase transient corrected pulses, the error per gate is reduced to $1.72\pm0.25\%$. However, the non-exponential behaviour of the decay curve is still observed. Note that this non-exponential behaviour was seen in the simulated RB results when $T_2^{*}$ is included (see the lower two curves in Fig. \ref{fig:bmsim}). This behaviour can be explained by the $T_2^{*}$ field distribution over the spin ensemble, which is a form of incoherent error and leads to a sum of exponential decays (see Appendix). Note that this type of incoherent error is not restricted to spatial ensemble systems, but also applies to single quantum systems when a parameter (e.g. local field) fluctuates between the many experiments that must be averaged to measure an observable~\cite{colmbm, lownoise}. After incorporating the selection sequence to effectively increase $T_2^{*}$ as described previously, the non-exponential behaviour is reduced greatly and the  decay curve fits well to a single exponential curve up to the largest number of gates (128) performed.

With the selection sequence, together with the longer unblanking delay and PTC pulses, the error per gate is reduced to $0.81\pm 0.04\%$ ($\square$ symbol in Fig. \ref{fig:bmresults}). This is nearly an order of magnitude improvement compared to the initial result of $6.7\%$. Simulation of the RB protocol including the selection sequence, and using experimentally determined values of $T_1$, $T_2$, $T_2^*$ and the $\bm{B}_1$ distribution, is also shown in Fig.~\ref{fig:bmresults} ($\circ$ symbol). The simulated error per gate is $0.76\pm0.03\%$, which is in good agreement with the experimental value. This agreement suggests that the error due pulse distortions has been reduced to the same order of magnitude as the error from intrinsic noise. Moreover, the improvement achieved by incorporating the selection sequence implies $T_2^*$ effect is a dominant error source. Therefore, although the standard RB protocol is not designed for explicitly distinguishing error sources, we are able to infer which mechanisms are dominant by comparison with simulation and by exploiting the selection sequence. 

\section{Conclusions and Future Work}
We constructed an X-band pulsed ESR spectrometer based on arbitrary waveform generation, a special loop-gap resonator and a pickup coil for monitoring the microwave field close to the sample. The RB protocol for the single qubit Clifford gates was used to characterize the precision of control over electron spin dynamics. The initial results indicated a relatively poor $93\%$ average gate fidelity. Active correction of phase transients together with stabilization of the microwave amplifier improved the gate fidelities significantly. By additionally selecting spin packets with a narrowed distribution of static fields, an average fidelity of $99.2\%$ was achieved. Comparison between experiment and simulations suggest that the remaining errors due to pulse imperfections are no larger than the intrinsic noise, i.e. noise due to inhomogeneous dephasing. In a sample with longer $T_2^*$, $T_2$ and $T_1$ values, the extrinsic errors due to microwave imperfections could again dominate. We expect that a suitable solid-state microwave amplifier should outperform the TWT amplifier in both noise figure and amplitude/phase stability, and it would be interesting to use RB for a side-by-side comparison of these amplifier technologies in the context of quantum control. RB protocols that can separate unitary control errors from incoherent errors due to  $T_2^*$, $T_2$ and $T_1$ effects have recently been developed \cite{coherence}, and can further distinguish sources of quantum gate error in future work. 

The RB of single qubit control performed here is a milestone towards more complex multi-qubit ESR QIP experiments. In future work, we are interested in characterizing microwave control during cryogenic operation, as high electron spin polarization is desirable. RB can also be used to characterize control in multi-qubit systems, e.g. an electron spin coupled to one or more nuclear spins by the hyperfine interaction~\cite{khaneja2007PRA,hodges2008universal,zhang2011coherent,simmons2011entanglement,filidou2012ultrafast}. High fidelity control in these systems can be achieved with optimal control pulses, so a clear next step is to perform RB on both single qubit and multi-qubit Clifford gates based on GRAPE \cite{khaneja2005optimal} pulses. If sufficiently high fidelities are achieved, then such hybrid electron-nuclear spin systems could test quantum algorithms such as heat bath algorithmic cooling \cite{HBAC_daniel,PPA,baugh2005experimental,RyanMultiAC} and quantum error correction \cite{knill1997theory,knill1998resilient,preskill1998reliable,knill2005quantum,aliferis2007accuracy,gottesman1997stabilizer}, in particular, multiple-round quantum error correction which is of fundamental importance for building a large scale quantum computer.

\section*{Acknowledgements}
This research was supported by NSERC, the Canada Foundation for Innovation, CIFAR, the province of Ontario, Industry Canada and the Gerald Schwartz and Heather Reisman Foundation. We thank David Cory and Troy Borneman for providing the sample and for stimulating discussions; Colm Ryan, Yingjie Zhang and Jeremy Chamilliard for their contributions to the spectrometer; Roberto Romero and Hiruy Haile for assistance with machining. 
\appendix
\section{Effect of incoherent error on the fidelity decay}
\label{sec:incoherent}
Incoherent error of a quantum process can be caused by classical noise, for example, a distribution over external experimental parameters~\cite{incoherent1,incoherent2} such as $\bm{B}_1$ inhomogeneity and $T_2^*$ line-broadening. In our system, after optimizing the TWT unblanking delay and using the phase transient corrected pulses, $T_2^{*}$ effect remains as the dominant source of the incoherent error. $\bm{B}_1$ inhomogeneity is also present, but does not critically reduce the control fidelity at current level. Hence only the $T_2^*$ noise is considered in our incoherent error discussions. In this section, we adapt the analysis presented in~\cite{colmbm} that was used to describe $\bm{B}_1$ inhomogeneity effect on the fidelity decay to explain how $T_2^{*}$ gives rise to the non-exponential decay observed in our experimental (Fig.~\ref{fig:bmresults}) and simulated (Fig.~\ref{fig:bmsim}) results.

Due to $T_2^*$ local field inhomogeneity across the sample, unitary errors with different strengths arise on the spins experiencing off-resonance fields. Intuitively, since the spins at different Larmor frequencies experience different average unitary errors, the signal decay curve should contain multiple depolarization rates, which explains the non-exponential behavior. For more concrete analysis, we consider a single step which consists of a computational gate ($S$) followed by a Pauli gate ($P$) in a randomized benchmarking sequence. The superoperator describing the process can be expressed as:
\begin{align}
\hat{\Lambda}=\int d\epsilon g(\epsilon)\hat{\Lambda}_{\epsilon}PS.\label{eq:singlePS}
\end{align}
Here $\epsilon$ is off-resonance frequency, $\hat{\Lambda}_{\epsilon}$ is the superoperator describing the cumulative error of $PS$ for the fraction of the system with off-resonance frequency $\epsilon$, and $g(\epsilon)$ is the distribution of $\epsilon$. $g(\epsilon)$ can be obtained from a frequency domain thermal state spectrum (Fig.~\ref{fig:t2star} for example). For a given $\epsilon$, the cumulative error strength of this single step can be calculated by
\begin{equation}
\label{eq:inc_error}
\xi=1-\frac{1}{4}|\text{Tr}(SPU_{inh}^{\dag})|^2,
\end{equation}
where $U_{inh}$ is the faulty implementation of $S$ and $P$ due to the local field inhomogeneity, and $\frac{1}{4}|\text{Tr}(SPU_{inh}^{\dag})|^2$ is the gate fidelity (Hilbert-Schmidt (HS) norm) of the faulty implementation. As the off-resonance fields cause undesired rotations along z-axis, which don't commute with rotations along x-axis or y-axis, the error strength $\xi$ is different for different combinations of $S$ and $P$. It can be easily verifed that for a given $\epsilon$ there are 9 different error strengths depending on different cases of $S$ and $P$ (e.g. whether $S$ and $P$ are rotations along z-axis, they are along parallel or anti-parallel axes, or the rotation axis of $S$ is clockwise or counterclockwise from that of $P$). The 9 types are labelled from 1 to 9 in Tab.~\ref{tab:errortype}.

\begin{table}[h!]
\centering
\begin{tabular}{  c |K{0.42cm}| K{0.73cm}|K{0.42cm}| p{0.75cm}|K{0.42cm}|K{0.69cm}|}
\cline{1-7}
\multicolumn{1}{  |c|}{\backslashbox{$P$}{$S$}} &\multicolumn{1}{K{0.42cm}|}{X90} &\multicolumn{1}{ K{0.73cm}|}{-X90}&\multicolumn{1}{ K{0.42cm}|}{Y90} &\multicolumn{1}{ K{0.73cm}|}{-Y90} &\multicolumn{1}{ K{0.42cm}|}{Z90} &\multicolumn{1}{ K{0.69cm}|}{-Z90} \\ \cline{1-7}
 \multicolumn{1}{ |c|}{X180} &\multicolumn{1}{ K{0.42cm}|}{1} &\multicolumn{1}{ K{0.73cm}|}{3}&\multicolumn{1}{ K{0.42cm}|}{4} &\multicolumn{1}{ K{0.73cm}|}{2} &\multicolumn{1}{ K{0.42cm}|}{7} &\multicolumn{1}{ K{0.69cm}|}{7} \\ \cline{1-7}
 \multicolumn{1}{ |c|}{-X180} &\multicolumn{1}{ K{0.42cm}|}{3} &\multicolumn{1}{ K{0.73cm}|}{1}&\multicolumn{1}{ K{0.42cm}|}{2} &\multicolumn{1}{ K{0.73cm}|}{4} &\multicolumn{1}{ K{0.42cm}|}{7} &\multicolumn{1}{ K{0.69cm}|}{7} \\ \cline{1-7}
 \multicolumn{1}{ |c|}{Y180} &\multicolumn{1}{ K{0.42cm}|}{2} &\multicolumn{1}{ K{0.73cm}|}{4}&\multicolumn{1}{ K{0.42cm}|}{1} &\multicolumn{1}{ K{0.73cm}|}{3} &\multicolumn{1}{ K{0.42cm}|}{7} &\multicolumn{1}{ K{0.73cm}|}{7} \\ \cline{1-7}
 \multicolumn{1}{ |c|}{-Y180} &\multicolumn{1}{ K{0.42cm}|}{4} &\multicolumn{1}{ K{0.73cm}|}{2}&\multicolumn{1}{ K{0.42cm}|}{3} &\multicolumn{1}{ K{0.73cm}|}{1} &\multicolumn{1}{ K{0.42cm}|}{7} &\multicolumn{1}{ K{0.69cm}|}{7} \\ \cline{1-7}
 \multicolumn{1}{ |c|}{Z180} &\multicolumn{1}{ K{0.42cm}|}{6} &\multicolumn{1}{ K{0.73cm}|}{6}&\multicolumn{1}{ K{0.42cm}|}{6} &\multicolumn{1}{ K{0.73cm}|}{6} &\multicolumn{1}{ K{0.42cm}|}{8} &\multicolumn{1}{ K{0.69cm}|}{8} \\ \cline{1-7}
 \multicolumn{1}{ |c|}{I} &\multicolumn{1}{ K{0.42cm}|}{5} &\multicolumn{1}{ K{0.73cm}|}{5}&\multicolumn{1}{ K{0.42cm}|}{5} &\multicolumn{1}{ K{0.73cm}|}{5} &\multicolumn{1}{ K{0.42cm}|}{9} &\multicolumn{1}{ K{0.69cm}|}{9} \\ \cline{1-7}
\end{tabular}

\caption[Gate-dependence of cumulative error due to $T_2^{*}$]{Gate-dependent cumulative error types for different combinations of computational gates $S$ (columns) and Pauli gates $P$ (rows) labelled from 1 to 9. The errors are grouped to 9 types according to their strengths defined in Eq.~\ref{eq:inc_error}. It should be mentioned that in our implementation rotations around z-axis are realized by changing the rotating frame definition, i.e. changing the phases of subsequent pulses and potentially the observation. In this way, a 180$^\circ$ rotation and a -180$^\circ$ rotation around  z-axis are the same. Therefore only Z180, which means a 180$^\circ$ rotation around  z-axis, is listed here.}
\label{tab:errortype}
\end{table}

\begin{figure}[h!]
\centering
\epsfig{file=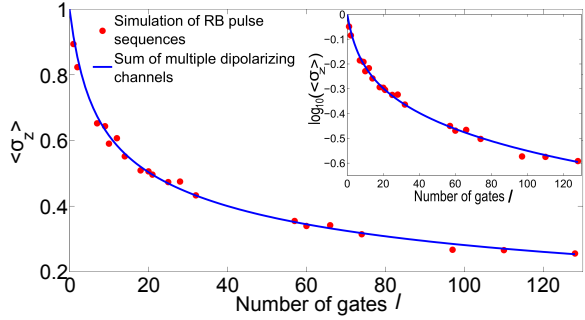,angle=0,width=0.5\textwidth}
\caption[Incoherent error analysis]{\label{fig:compare}Comparison between the numerically calculated prediction using Eq.~\ref{eq:ngate} (solid) and the realistic simulation averaged over 420 randomized benchmarking pulse sequences ($\bullet$). Excellent agreement between the two is observed. $T_1$ and $T_2$ processes are not taken into account.}
\end{figure}

Therefore, for a fixed $\epsilon$, the error is gate-dependent. However, it is proven in~\cite{emersonRBM,PhysRevLett.106.180504,PhysRevA.85.042311} that the cumulative effect of the gate-dependent errors in the randomized benchmarking can still be described as a depolarizing channel as long as the gate dependence is weak. 
Therefore, for the spin packet with the off-resonance frequency $\epsilon$, upon averaging over random gate sequences, $\hat{\Lambda}_{\epsilon}$ in Eq.~\ref{eq:singlePS} forms a depolarizing channel $\hat{\Lambda}_{\epsilon,ave}$ with the depolarizing factor $p_{\epsilon}$. $\hat{\Lambda}_{\epsilon,ave}$ and $p_{\epsilon}$ can be expressed as~\cite{colmbm}:
\begin{align}
\hat{\Lambda}_{\epsilon,ave}&=\sum_{i=1}^9w_{i}\hat{\Lambda}_{\epsilon,ave,i},\\
p_{\epsilon}&=\sum_{i=1}^9w_{i}p_i.
\end{align}
Here $\hat{\Lambda}_{\epsilon,ave,i}$ is the depolarized channel associated with the error type $i$ with depolarizing parameter $p_i=(4-|\text{Tr}(SPU_{inh}^{\dagger})|^2)/3=4\xi_i/3$~\cite{emersonRBM}. $w_i$ is the probability for the error type $i$ to occur, and their values are 1/9, 1/9, 1/9, 1/9, 1/9, 1/9, 2/9, 1/18, 1/18. 

Finally, by averaging over the distribution of $\epsilon$, the expression for the channel constructed from $n$ random gates can be obtained as:
\begin{align}
\hat{\Lambda}_{ave}(n)=\int d\epsilon g(\epsilon)\hat{\Lambda}_{\epsilon, ave}^n,\label{eq:ngate}
\end{align}
where $\hat{\Lambda}_{\epsilon, ave}^n(\rho)=(1-p_{\epsilon})^n\rho$ and $\rho$ is the deviation density matrix or the traceless part of the full density matrix. Therefore, the fidelity decay (in our case, the decay of the expectation value of $\sigma_z$) is the sum of multiple exponential decays weighted with the distribution function $g(\epsilon)$.

Using experimentally measured values for $g(\epsilon)$, we compare in Fig.~\ref{fig:compare} the decay curve numerically calculated from Eq.~\ref{eq:ngate} with the curve obtained by simulating randomized benchmarking sequences. The evaluation of Eq.~\ref{eq:ngate} is carried out numerically since $U_{inh}$ is generated by a time-dependent Hamiltonian with two non-commuting terms, $\sigma_z$ term representing the off-resonance effect and $\sigma_{x,y}$ terms for the external control. For the $S$ and $P$ gates, 35 ns Gaussian pulses with 1 ns time steps are used. Two decay curves agree very well, indicating the $T_2^*$ incoherent error is responsible for the non-exponential behavior of the fidelity decay. Furthermore, in the experimental data, the decay curve fits very well to a single-exponential function within the error bar when $T_2^{*}$ is extended using the selection sequence ($\square$ in Fig.~\ref{fig:bmresults}). Therefore, we conclude that the non-exponential fidelity decay can be explained by the $T_2^{*}$ error.





\end{document}